\documentclass[a4paper,showkeys,aps,pre,preprint,showpacs,preprintnumbers,amsmath,amssymb]{revtex4}
\usepackage{amsthm}
\usepackage[german,english]{babel}
\usepackage[unicode]{hyperref}

\begin{document}
	
\title{Radiative Transfer: Asymptotic Solutions of the Kinetic Equation of Radiation Propagation, $n$th Order Asymptotic Approximation and Improved Boundary Conditions}
	
	\author{S.~A.~Serov}
	\email{S.A.Serov@inbox.ru}
	\affiliation{
		Institute of Theoretical and Mathematical Physics, 
		Sarov, Russia}
	\author{S.~S.~Serova}
	\email{S.S.Serova@inbox.ru}
	\affiliation{St. Petersburg State University, 
		St. Petersburg, Russia}
	
	\date{\today}
	
	\begin{abstract}
		In the article, new asymptotic approximation of the $n$th order is obtained and proposed to be used in calculations of radiation propagation without scattering in optically thick media; the asymptotic approximation is much simpler and more precise than the known diffusion approximation. 
		The rigorous derivation of the diffusion approximation equation and the equation of the radiation heat conduction approximation is obtained from the constructed asymptotic solution of the kinetic equation of radiation propagation in optically thick media.
		It is shown, that for optically thick media the asymptotic solution of the kinetic equation of radiation propagation without scattering is asymptotic expansion of the exact integral solution of that kinetic equation.
		Improved boundary conditions, which are essential for practical application in calculations of radiation propagation, are derived (for inner boundaries and outer boundaries with vacuum). 
	\end{abstract}
	
	
	
	\maketitle

	\section{Introduction}
	\label{sec:Introduction}
	
In the article, radiation propagation in non-scattering media under the assumption of \textit{local thermodynamic equilibrium} is discussed [the temperature and density of the substance are assumed to be locally determined, since the kinetic equation of the radiation propagation \eqref{kin_eq} (considered below) includes the radiation absorption coefficient $ \kappa ^{\prime}_{\nu } $, which is reduced by induced emission 
and depends on the properties of the substance, its density and temperature, and the spectral intensity of the equilibrium radiation $I_{\nu \text{P}} $ depending only on the temperature].

According to the estimates, radiation scattering is significant in comparison to radiation absorption in a very 
rarefied and fully ionized gas \textit{only} when the bremsstrahlung absorption, which is proportional to the product of the ion number density of the substance and the free electron number density $N_{+}N_{e}$, becomes small --- \cite[Chapter II, \S~2]{Zeldovich2012} (see also, for example, \cite[Chapter 4, \S\S~4.4--4.5]{Mihalas1978stellar}).
Therefore, radiation scattering can be neglected in comparison to radiation absorption, when describing physical processes inside stars, when solving problems of controlled thermonuclear fusion and other problems of high-energy physics.
In this case, the kinetic approximation (the kinetic equation of radiation propagation is solved numerically), the diffusion approximation and the radiation heat conduction approximation (see, for example, \cite{Zeldovich2012}, \cite{Mihalas1978stellar}) are currently used in the calculations of radiation propagation (gas-dynamic calculations with radiation).
Performance of calculations using kinetic approximation, which is the most precise from a physical point of view, is significantly more time-consumed.
It is difficult to use kinetic approximation in calculations of radiation propagation in optically very thick media (for example, in substances with a high density and a big charge of an atomic nucleus when mean free path of radiation is very small). 
In addition, the natural difference approximation of first order derivatives in the kinetic equation of radiation propagation leads to non-monotonic difference schemes.
Because of it, most part of calculations of radiation propagation in optically thick media without scattering are still carried out in the radiation heat conduction approximation and the diffusion approximation.
At the same time, the radiation heat conduction approximation and the diffusion approximation known so far have the zero order of the asymptotic precision (see below), therefore, badly describe angular distribution of radiation in optically not very thick media (use of these approximations in optically thin environments is not justified).

In the Section \ref{sec:asympt_solution}, asymptotic solutions of the kinetic equation of radiation propagation without scattering are considered in two limiting cases: optically thick and optically thin media.
A system of equations of the \textit{$n$th order asymptotic approximation} (integer $n \ge 1$), which is proposed to be used in calculations of radiation propagation in optically thick media instead of the diffusion approximation and the radiation heat conduction approximation, is obtained from the asymptotic solution of the kinetic equation of radiation propagation for optically thick media.
In calculations of radiation propagation in optically thick media at critical points, it is proposed to use limitations on radiation mean free paths for the radiation heat conduction approximation as well as for the asymptotic and diffusion approximations.
In particular, \textit{rigorous justification} of the coefficient 
${1\mathord{\left/ {\vphantom {1 3}} \right. \kern-\nulldelimiterspace} 3}$ 
in the second diffusion approximation equation and in the radiant energy flux equation of the radiation heat conduction approximation follows from the constructed asymptotic solution of the kinetic equation of radiation propagation for optically thick media without scattering.

From the theoretical point of view, considered kinetic equation of radiation propagation in media without scattering is interesting, for its exact integral solution is known.
In the Section \ref {sec:integral_solution}, 
the asymptotic solution of the kinetic equation of radiation propagation for optically thick media is shown to be different from the exact integral solution only in the residual, which is small for optically thick media, 
i.e. for optically thick media, the asymptotic solution of the kinetic equation of radiation propagation without scattering is \textit{asymptotic expansion} of the exact integral solution of this kinetic equation,
and the uncommon terms containing time derivatives correspond to the ``retarded'' radiation.

When the new asymptotic approximation of the $n$th order or the diffusion approximation and the radiation heat conduction approximation of increased precision are used in calculations of radiation propagation, boundary conditions of higher order of 
the asymptotic precision shall be used for the general accuracy of calculations not to reduce.
Therefore, in the Section \ref{sec:b_condition}, we obtain the \textit{improved boundary conditions} (for internal boundaries and external boundaries with vacuum) from the asymptotic solution of the kinetic equation of radiation propagation for optically thick non-scattering media, which are proposed to be used in calculations of radiation propagation.
In the expressions for the improved boundary conditions, the limitations on radiation mean free paths are introduced, similar to those considered in the Section \ref{sec:asympt_solution}.


Asymptotic solutions of the kinetic equation of radiation propagation without scattering were considered in \cite{Larsen1983} as a special case (to take into account the possible radiation scattering in a very specific asymptotic limit).
For asymptotic solutions of the kinetic equation of radiation propagation with scattering, see, for example, 
articles \cite{habetler1975uniform}, \cite{larsen1976asymptotic}, \cite{larsen1977asymptotic}, \cite{larsen2010asymptotic}. 
For approximations with a limitation of the radiant energy flux, see, for example, the review \cite{Brunner2002forms} and the references therein.

The notations used in the article are close to the notations in \cite{Zeldovich2012}, the obvious dependences of the functions under consideration (on time, spatial coordinates, and directions \ldots ) are omitted.
Relativistic effects are not considered below.
Unless otherwise stated, it is assumed that a rectangular Cartesian coordinate system $ \left( x,y,z \right) = \left( x^1,x^2,x^3 \right) $  is introduced in three-dimensional space.
The indices of three-dimensional vectors and tensors are denoted by Greek letters and range from 1 to 3; the summation over the same upper and lower coordinate indices of three-dimensional tensors is implied; if some indices are not coordinate tensor indices, then the sum sign is indicated explicitly, if necessary.
The metric is considered to be determined by the diagonal metric tensor, which components are:
$g_{\alpha \beta}=0$ if $\alpha  \ne \beta$, $g_{11}=g_{22}=g_{33}=1$; the components of the radiation pressure tensor below are
written through components $ g^{\alpha \beta} $ of the contravariant metric tensor,
which are determined by the equation $ g_{\alpha \beta}\, g^{\beta \xi }= {\delta}_{\alpha}^{\xi}$, where $ {\delta}_{\alpha}^{\xi} $ --- 
$ {\delta} $-Kronecker symbol of the rank 2 (components of the unit three-dimensional tensor), ${\delta}_{\alpha}^{\xi}=0$ if 
$\alpha  \ne {\xi}$, ${\delta}_{\alpha}^{\xi}=1$ if $\alpha = {\xi}$; $g^{\alpha \beta}=0$ if $\alpha  \ne \beta$, $g^{11}=g^{22}=g^{33}=1$; for four-dimensional coordinates are also used the notations $t$ --- time, ${\mathbf{r}}$ --- three-dimensional radius-vector.

\section{Asymptotic solutions of the kinetic equation of radiation propagation}
\label{sec:asympt_solution}

If radiation scattering is neglected in the general kinetic equation of radiation propagation \{see, for example, the equation (2.24) in \cite{Zeldovich2012}, the equation (2.26) in \cite{Mihalas1978stellar}; cf. the the equation (2.26) in \cite{Ambartsumian1958} or the equation (48) in \cite[Chapter~I]{chandrasekhar1960radiative}\} and the emission coefficient is expressed through the absorption coefficient and the spectral intensity of the equilibrium radiation in the state of thermodynamic equilibrium under the assumption of local thermodynamic equilibrium in accordance with \textit{Kirchhoff's law} \{see the equations (2.21) and (2.27) in \cite{Zeldovich2012} or the equations (2.4) and (2.5) in \cite{Mihalas1978stellar}\}, the approximate kinetic equation of radiation propagation without scattering, which is a first order partial differential equation for the spectral intensity of the radiation 
$I_{\nu }\left(t,\mathbf{r},\mathbf{\Omega }\right)$ (depending on time, spatial coordinates and direction), can be written in the following form (cf. the equation (2.28) in \cite{Zeldovich2012}):
\begin{eqnarray} \label{kin_eq} 
\left(\frac{\partial }{c\partial t} +\mathbf{\Omega }\cdot \nabla \right)I_{\nu } =\kappa ^{\prime}_{\nu } \left(I_{\nu \text{P}} -I_{\nu } \right)\, , 
\end{eqnarray} 
In the equation \eqref{kin_eq}, $\nu $ --- frequency of photons; $c$ --– speed of light in vacuum; 
\begin{eqnarray} \label{I_nu} 
I_{\nu } \left(t,\mathbf{r},\mathbf{\Omega }\right)d\nu d\mathbf{\Omega }
=h\nu cf\left( {\nu ,t,{\mathbf{r}},{\mathbf{\Omega }}} \right) d\nu d\mathbf{\Omega }
\end{eqnarray} 
gives the amount of radiant energy (per unit time through the unit area placed at the point $\mathbf{r}$, perpendicular to direction 
$\mathbf{\Omega }$) in the spectral interval $d\nu $, which is transferred by light quanta with energy $ h \nu $ that have a motion direction in the solid angle element $d\mathbf{\Omega }$ around the unit vector $\mathbf{\Omega }$; 
$h$ --- Planck constant; 
$ f\left( {\nu ,t,{\mathbf{r}},{\mathbf{\Omega }}}  \right)$ --- number density of light quanta (which is similar to the velocity distribution function of the gas particles in the kinetic theory of rarefied gases) in the spectral interval from $ \nu $ to $ \nu + d\nu $, which are located at the moment $ t $ near the point $ {\mathbf{r}} $ and have a motion direction in the solid angle element $ d\mathbf{\Omega } $ around the unit vector $ \mathbf{\Omega } $; 
$I_{\nu \text{P}} $ --- the spectral intensity of the equilibrium radiation in the state of thermodynamic equilibrium, determined by Planck formula:
\begin{eqnarray} \label{I_nuP} 
I_{\nu \text{P}} \equiv B_{\nu } \equiv \frac{2h\nu ^{3} }{c^{2} } \frac{1}{e^{\frac{h\nu }{kT} } -1} \; ; 
\end{eqnarray} 
$k$ --- Boltzmann constant; $T$ --- temperature of a substance, measured on a thermodynamic temperature scale; 
$\kappa ^{\prime}_{\nu } \equiv {1\mathord{\left/ {\vphantom {1 l^{\prime}_{\nu } }} \right. \kern-\nulldelimiterspace} l^{\prime}_{\nu } } $ --- radiation absorption coefficient reduced by induced emission
(see, for example,  \cite[Chapter~II, \S~5]{Zeldovich2012})
\begin{eqnarray} \label{3} 
\kappa ^{\prime}_{\nu } \equiv \kappa _{\nu } \left(1-e^{-\frac{h\nu }{kT} } \right)\, ; 
\end{eqnarray} 
$\kappa _{\nu } I_{\nu } d\nu d\mathbf{\Omega }$ is equal to the energy of light quanta with a frequency from $\nu $ to $\nu +d\nu $, which have the motion direction in the solid angle element  $d\mathbf{\Omega }$ around the unit vector $\mathbf{\Omega }$ and are absorbed by the substance per unit volume per unit time;
$l^{\prime}_{\nu } $ --- mean free path of photon with the frequency $\nu $.

Let us construct asymptotic solutions of the equation \eqref{kin_eq} for two limiting cases: ${l^{\prime}_{\nu } \mathord{\left/ {\vphantom {l^{\prime}_{\nu }  L\to 0}} \right. \kern-\nulldelimiterspace} L\to 0} $ (optically thick media) and ${L \mathord{\left/ {\vphantom {L  l^{\prime}_{\nu }\to 0}} \right. \kern-\nulldelimiterspace} l^{\prime}_{\nu }\to 0} $ (optically thin media), where $L$ --- characteristic distance of radiation intensity change.
In order not to change the form of a physical equation, it is convenient not to select an explicitly small parameter (however, in the cases under consideration, it is very simple to select a small parameter; it is sufficient to use dimensionless independent variables in the equation \eqref{kin_eq}:
$\left\{{ct\mathord{\left/ {\vphantom {ct L,\, \; {r\mathord{\left/ {\vphantom {r L}} \right. \kern-\nulldelimiterspace} L} }} \right. \kern-\nulldelimiterspace} L,\, \; {r\mathord{\left/ {\vphantom {r L}} \right. \kern-\nulldelimiterspace} L} } \right\}$),  in the powers of which an \textit{asymptotic expansion with variable coefficients} is constructed (see \cite[Chapter~V, 
\S ~2, Section~5]{bourbaki2004}), when it tends to zero, but to introduce this small parameter into a physical equation as an indicator of the smallness of the corresponding terms of a physical equation.

For optically thick media, we introduce a small parameter $\varepsilon $ in the equation \eqref{kin_eq} as follows:
\begin{eqnarray} \label{eps_thick} 
\varepsilon l^{\prime}_{\nu } \left(\frac{\partial }{c\partial t} +\mathbf{\Omega }\cdot \nabla \right)I_{\nu } =\left(I_{\nu \text{P}} -I_{\nu } \right)\, , 
\end{eqnarray} 
\begin{eqnarray} \label{l_L_to_0} 
\frac{l^{\prime}_{\nu } }{L} \to 0\, . 
\end{eqnarray} 
Let us write the asymptotic expansion of the spectral radiation intensity in the form of a formal series of successive approximations in powers of $\varepsilon $:
\begin{eqnarray} \label{exp_I_thick} 
I_{\nu } =\varepsilon ^{0} I_{\nu }^{\left(0\right)} +\varepsilon I_{\nu }^{\left(1\right)} +\varepsilon ^{2} I_{\nu }^{\left(2\right)} +\cdots \, \, , 
\end{eqnarray} 
and substitute this power series into the equation \eqref{eps_thick} and equate the variable coefficients with the same powers $\varepsilon $.

As a result, we obtain the following system of equations of the method of successive approximations:
\begin{subequations}
	\label{exp_I_thick_01n}
	\begin{eqnarray}
	I_{\nu }^{\left(0\right)}   &=& I_{\nu \text{P}} ,
	\label{I_0}
	\\
	I_{\nu }^{\left(1\right)}   &=& -l^{\prime}_{\nu } \left(\frac{\partial }{c\partial t} +\mathbf{\Omega }\cdot \nabla \right)I_{\nu \text{P}} ,
	\label{I_1}
	\\
	&{\vdots }& \nonumber\\
	I_{\nu }^{\left(n\right)}   &=& -l^{\prime}_{\nu } \left(\frac{\partial }{c\partial t} +\mathbf{\Omega }\cdot \nabla \right)I_{\nu }^{\left(n-1\right)} .
	\label{I_n}
	\end{eqnarray}
\end{subequations}

From the equations \eqref{exp_I_thick} and \eqref{exp_I_thick}, in accordance with the general definitions of the spectral radiant energy density
\begin{eqnarray} \label{def_U} 
U_{\nu } \; \mathop{=}\limits^{\text{def}} \; \frac{1}{c} \int _{\left(4\pi \right)}I_{\nu } d\mathbf{\Omega }  
\end{eqnarray} 
and the spectral radiant energy flux
\begin{eqnarray} \label{def_S} 
\mathbf{S}_{\nu } \; \mathop{=}\limits^{\text{def}} \; \int _{\left(4\pi \right)}I_{\nu } \mathbf{\Omega }\, d\mathbf{\Omega }  
\end{eqnarray} 
we obtain the asymptotic expansion of the spectral radiant energy density 
\begin{eqnarray} \label{exp_U_thick} 
U_{\nu } =\varepsilon ^{0} U_{\nu }^{\left(0\right)} +\varepsilon U_{\nu }^{\left(1\right)} +\varepsilon ^{2} U_{\nu }^{\left(2\right)} +\cdots , 
\end{eqnarray} 
where, in particular,
\begin{eqnarray} \label{U0_nu} 
U_{\nu }^{\left(0\right)} =\frac{4\pi }{c} I_{\nu \text{P}} = U_{\nu \text{P}} 
\end{eqnarray} 
--- equilibrium spectral radiant energy density,
\begin{eqnarray} \label{U1_nu} 
U_{\nu }^{\left(1\right)} =-\, l^{\prime}_{\nu } \frac{\partial U_{\nu \text{P}} }{c\partial t} , 
\end{eqnarray} 
\begin{eqnarray} \label{U2_nu} 
U_{\nu }^{\left(2\right)} =\, l^{\prime}_{\nu } \frac{\partial  }{c\partial t}\left(l^{\prime}_{\nu } \frac{\partial U_{\nu \text{P}} }{c\partial t}\right)
+ \frac{1}{3}\, l^{\prime}_{\nu } {\nabla \cdot}\left(l^{\prime}_{\nu } \nabla U_{\nu \text{P}}\right), 
\end{eqnarray} 
and the asymptotic expansion of the spectral radiant energy flux
\begin{eqnarray} \label{exp_S_thick} 
\mathbf{S}_{\nu } =\varepsilon ^{0} \mathbf{S}_{\nu }^{\left(0\right)} +\varepsilon \mathbf{S}_{\nu }^{\left(1\right)} +\varepsilon ^{2} \mathbf{S}_{\nu }^{\left(2\right)} +\cdots , 
\end{eqnarray} 
where
\begin{eqnarray} \label{S_0)} 
\mathbf{S}_{\nu }^{\left(0\right)} =0, 
\end{eqnarray} 
\begin{eqnarray} \label{S_1} 
\mathbf{S}_{\nu }^{\left(1\right)} =-\frac{c}{3} \, l^{\prime}_{\nu } \nabla U_{\nu \text{P}} =-\frac{c}{3} l^{\prime}_{\nu } \nabla U_{\nu }^{\left(0\right)}  , 
\end{eqnarray} 
\begin{eqnarray} \label{S_2} 
\mathbf{S}_{\nu }^{\left(2\right)} 
 &=& \frac{c}{3} 
\, l^{\prime}_{\nu } \nabla \left( l^{\prime}_{\nu } \frac{\partial U_{\nu \text{P}}}{c\partial t} \right) + \frac{c}{3} 
\, l^{\prime}_{\nu } \frac{\partial  }{c\partial t}\, \left(l^{\prime}_{\nu } \nabla U_{\nu \text{P}}\right)\nonumber\\
 &=& - \frac{c}{3} 
\, l^{\prime}_{\nu } \nabla U_{\nu }^{\left(1\right)} - 
\, l^{\prime}_{\nu } \frac{\partial \mathbf{S}_{\nu }^{\left(1\right)} }{c\partial t} . 
\end{eqnarray}

The spectral pressure tensor of radiation is defined similarly to the pressure tensor of gas particles in the kinetic theory of rarefied gases
\begin{eqnarray} \label{def_P_nu} 
{\text{p}}_{\nu }
\; \mathop{=}\limits^{\text{def}} \; \frac{1}{c} \int _{\left(4\pi \right)} I_{\nu }\, \mathbf{\Omega }  \otimes \mathbf{\Omega }\, d\mathbf{\Omega } ,
\end{eqnarray} 
or in the components
\begin{eqnarray} \label{def_P_nu_ab} 
{\text{p}}_{\nu }^{\alpha \beta }
\; \mathop{=}\limits^{\text{def}} \; \frac{1}{c} 
\int _{\left(4\pi \right)} I_{\nu }\, {\Omega }^{\alpha } {\Omega }^{\beta }\, d\mathbf{\Omega } 
=\int _{\left(4\pi \right)} \left[ f\left( {\nu ,t,{\mathbf{r}},{\mathbf{\Omega }}} \right) c \, {\Omega }^{\alpha } \right] \left( \frac{h\nu }{c}\,{\Omega }^{\beta } \right) d\mathbf{\Omega } 
\end{eqnarray} 
--- see, for example, the equations (66.1)-(66.3) in \cite{mihalas1995foundations}.

We obtain the asymptotic expansions of the components of the spectral pressure tensor of radiation from the equations \eqref{exp_I_thick}-\eqref{exp_I_thick_01n} and \eqref{def_P_nu_ab}
\begin{eqnarray} \label{exp_P_nu_ab} 
{\text{p}}_{\nu }^{\alpha \beta } =\varepsilon ^{0}\, {\text{p}}_{\nu }^{\alpha \beta \left(0\right)} +\varepsilon \, {\text{p}}_{\nu }^{\alpha \beta \left(1\right)} +\varepsilon ^{2}\, {\text{p}}_{\nu }^{\alpha \beta \left(2\right)} +\cdots , 
\end{eqnarray} 
where 
\begin{eqnarray} \label{P_nu_ab0} 
{\text{p}}_{\nu }^{\alpha \beta \left(0\right)} 
= \frac{1}{3} U_{\nu \text{P}} {g^{\alpha \beta}}
= \frac{1}{3} U_{\nu }^{\left(0\right)} {g^{\alpha \beta}}, 
\end{eqnarray} 
\begin{eqnarray} \label{P_nu_ab1} 
{\text{p}}_{\nu }^{\alpha \beta \left(1\right)} 
=\frac{1}{3} \left( - l^{\prime}_{\nu } \frac{\partial U_{\nu \text{P}}}{c\partial t} \right) {g^{\alpha \beta}}
= \frac{1}{3} U_{\nu }^{\left(1\right)} {g^{\alpha \beta}} , 
\end{eqnarray} 
\begin{eqnarray} \label{P_nu_ab2} 
{\text{p}}_{\nu }^{\alpha \beta \left(2\right)} 
 &=& \frac{1}{c}\, 
l^{\prime}_{\nu }\, \frac{\partial }{\partial x^{\xi }}\left( l^{\prime}_{\nu }\,  \frac{\partial I_{\nu \text{P}} }{\partial x^{\zeta }} \right) 
\int _{\left(4\pi \right)} {\Omega }^{\xi } {\Omega }^{\zeta } 
{\Omega }^{\alpha } {\Omega }^{\beta }\, d\mathbf{\Omega }
+ \frac{1}{3}\, l^{\prime}_{\nu } \frac{\partial  }{c\partial t}\left(l^{\prime}_{\nu } \frac{\partial U_{\nu \text{P}} }{c\partial t}\right) {g^{\alpha \beta}} \nonumber\\
 &=& \frac{1}{15}\, 
l^{\prime}_{\nu }\, \frac{\partial }{\partial x^{\xi }}\left( l^{\prime}_{\nu }\,  \frac{\partial U_{\nu \text{P}} }{\partial x^{\zeta }} \right) 
\left( 
g^{\xi \zeta }  g^{\alpha \beta } + 
g^{\xi \alpha } g^{\zeta \beta } +
g^{\xi \beta }  g^{\alpha \zeta }
\right)
\nonumber\\
 && + \frac{1}{3}\, l^{\prime}_{\nu } \frac{\partial  }{c\partial t}\left(l^{\prime}_{\nu } \frac{\partial U_{\nu \text{P}} }{c\partial t}\right) {g^{\alpha \beta}}  
\end{eqnarray} 
and
\begin{eqnarray} \label{int_4omega} 
\int _{\left(4\pi \right)} {\Omega }^{\xi } {\Omega }^{\zeta } 
{\Omega }^{\alpha } {\Omega }^{\beta }\, d\mathbf{\Omega }
= \frac{2}{\left( 2+1\right) \left( 2+2+1\right) }\, {2\pi }\left( 
g^{\xi \zeta }  g^{\alpha \beta } + 
g^{\xi \alpha } g^{\zeta \beta } +
g^{\xi \beta }  g^{\alpha \zeta }
\right) 
\end{eqnarray} 
(other integrals over the angles are calculated and written analogically).


The equations \eqref{exp_I_thick_01n}, \eqref{exp_U_thick}, \eqref{exp_S_thick}, \eqref{exp_P_nu_ab} form the system of equations of the \textit{$n$th order asymptotic approximation}, that can be used in calculations of radiation propagation (gas-dynamic calculations with radiation) instead of the equations of the diffusion approximation and the radiation heat conduction approximation, which are considered below.

For $ \varepsilon = 0 $ ($ l^{\prime}_{\nu } \equiv 0 $), the asymptotic solution \eqref{exp_I_thick} of the equation \eqref{eps_thick} is exact:
\begin{eqnarray}
I_{\nu } = I_{\nu }^{\left(0\right)} = I_{\nu \text{P}} ,
\label{I_eps0}
\end{eqnarray}
see the equation \eqref{I_0}: 
photons with anisotropic angular distribution (for example, near medium boundary) are absorbed on the zero piece of the way and isotropically re-emitted according to the spectral intensity of the equilibrium radiation, determined by Planck formula.

The spectral radiation intensity is isotropic in the zero order asymptotic approximation according to the equation \eqref{I_0}. Angular dependence of the radiation intensity appears only since the first order of the asymptotic precision --- see the equations \eqref{I_1}, \eqref {I_n}. 
Therefore, it is necessary to use, at least, the first order asymptotic approximation in gas-dynamic calculations with radiation for the correct description of angular distributions of gas-dynamic quantities depending on radiation, for example, temperature and pressure (this note belongs also to the diffusion approximation and to the radiation heat conduction approximation considered below).
The equations \eqref{exp_I_thick_01n} for spectral radiation intensity are not usually used as such in gas-dynamic calculations with radiation.

The asymptotic approximation of the $n$th order can be used in calculations of radiation propagation only if the series \eqref{exp_I_thick}, \eqref{exp_U_thick}, \eqref{exp_S_thick}, \eqref{exp_P_nu_ab} converge.
However convergence of these power series can be broken even in optically thick media at critical points, for example, near thermal (shock) waves fronts.
For ensuring convergence of the power series \eqref{exp_I_thick}, \eqref{exp_U_thick}, \eqref{exp_S_thick}, \eqref{exp_P_nu_ab} in calculations of radiation propagation, it is possible to use limitations on radiation mean free paths, which are written down separately for time and space derivatives:
\begin{eqnarray} \label{rep_l_t} 
l^{\prime}_{\nu } \frac{\partial }{c\partial t} \ \longrightarrow \  
\tilde{l}_{\nu }^{\, t} \frac{\partial }{{c\partial t}} , 
\end{eqnarray} 
\begin{eqnarray} \label{rep_l_s} 
l^{\prime}_{\nu }\, \nabla \ \longrightarrow \  
\tilde{l}_{\nu }^{\, s}\, \nabla , 
\end{eqnarray} 
where
\begin{eqnarray} \label{lim_l_t} 
{\tilde{l}_{\nu }^{\, t}} =\min \left\{{l_{\nu }^{\prime} } ,\, 
K_{\nu }^{t} {L_{\nu }^{t} } \right\} , 
\end{eqnarray} 
\begin{eqnarray} \label{lim_l_s} 
{\tilde{l}_{\nu }^{\, s}} =\min \left\{{l_{\nu }^{\prime} } ,\, 
K_{\nu }^{s} {L_{\nu }^{s} } \right\} , 
\end{eqnarray} 
\begin{eqnarray} \label{L_nu_t} 
{L_{\nu }^{t} } 
= \frac{I_{\nu \text{P}}}{ 
	{\left|{\partial I_{\nu \text{P}} \mathord{\left/ {\vphantom {\partial I_{\nu \text{P}}  c\partial t}} \right. \kern-\nulldelimiterspace} c\partial t} \right|}} 
= \frac{U_{\nu \text{P}}}{
	{\left|{\partial U_{\nu \text{P}} \mathord{\left/ {\vphantom {\partial U_{\nu \text{P}}  c\partial t}} \right. \kern-\nulldelimiterspace} c\partial t} \right|}} , 
\end{eqnarray} 
\begin{eqnarray} \label{L_nu_s} 
{L_{\nu }^{s} } 
= \frac{I_{\nu \text{P}}}{{\left|{\nabla I_{\nu \text{P}} \mathord{{\vphantom {\nabla I_{\nu \text{P}}}}  \kern-\nulldelimiterspace}} \right|} } 
= \frac{U_{\nu \text{P}}}{{\left|{\nabla U_{\nu \text{P}} \mathord{{\vphantom {\nabla U_{\nu \text{P}}}}  \kern-\nulldelimiterspace}} \right|}} , 
\end{eqnarray} 
\begin{eqnarray} \label{sum_K_nu} 
K_{\nu }^{t}+K_{\nu }^{s} \lesssim 1 , 
\end{eqnarray} 
cf. the equation \eqref{I_n}.
The limitations \eqref{rep_l_t} and \eqref{rep_l_s} (like other limitations on the radiant energy flux, which are currently used in the calculations of radiation propagation, see, for example, \cite{Brunner2002forms}) cannot be considered as theoretically justified, but it should be noted that they ``work'' only at critical calculating points for optically thick media.

The first equation of the diffusion approximation is obtained (exactly) from the kinetic equation of radiation propagation \eqref{kin_eq} by integrating the equation \eqref{kin_eq} over the angles:
\begin{eqnarray} \label{dif_approx_eq1} 
\frac{\partial U_{\nu } }{\partial t} 
+ \text{div}\, \mathbf{S}_{\nu } 
= c \kappa ^{\prime}_{\nu } \left(U_{\nu \text{P}} -U_{\nu } \right)\, . 
\end{eqnarray} 
The second diffusion approximation equation can be obtained from the equation \eqref{S_1} by adding higher-order terms of smallness of the asymptotic expansions of the spectral radiant energy density \eqref{exp_U_thick} and the spectral radiant energy flux \eqref{exp_S_thick} into the equation \eqref{S_1}:
\begin{eqnarray} \label{dif_approx_eq2} 
\mathbf{S}_{\nu } =-\frac{c}{3} l^{\prime}_{\nu } \nabla U_{\nu } \, . 
\end{eqnarray} 
Due to the presence of the second term in the right side of the equation \eqref{S_2}, the second diffusion approximation equation has the first order of the asymptotic precision (when $ \frac{l^{\prime}_{\nu } }{L} \to 0 $). 

The diffusion approximation equations \eqref{dif_approx_eq1}, \eqref{dif_approx_eq2} are a system of two first order partial differential equations for two unknown  functions of the spatial coordinates and time, that depend on the radiation frequency (but not on the direction): the spectral radiant energy density and the spectral radiant energy flux.
For optically thick media, the diffusion approximation is \underbar{not more precise} than the first order asymptotic approximation, but the first order asymptotic approximation is much simpler.
Gas-dynamic calculations with radiation
have the zero order of the asymptotic precision, if usual boundary conditions on radiation of the zero order of the asymptotic precision are used or if the expression for radiation pressure is used that follows from an equation of state of a substance with radiation --- cf. the equation \eqref{P_nu_ab0}.

In calculations of the radiation propagation in optically thick media at critical points, the limitations on radiation mean free paths \eqref{rep_l_t} and \eqref{rep_l_s} can be used.


In the radiation heat conduction approximation, the total (integrated over all frequencies) radiant energy flux is given by an analytical formula, see the equation (2.76) in \cite{Zeldovich2012}:
\begin{eqnarray} \label{S_RHC} 
\mathbf{S}=-\frac{c}{3} l^{\prime}_\text{R} \nabla U_{\text{P}} =-\frac{16l^{\prime}_\text{R} \sigma T^{3} }{3} \nabla T\, ; 
\end{eqnarray} 
in this equation
\begin{eqnarray} \label{31)} 
l^{\prime}_\text{R} =\frac{\int _{0}^{\infty }l^{\prime}_{\nu } \frac{dI_{\nu \text{P}} }{dT} d\nu  }{\int _{0}^{\infty }\frac{dI_{\nu \text{P}} }{dT} d\nu  } =\frac{\int _{0}^{\infty }\frac{1}{\kappa ^{\prime}_{\nu } } \frac{dI_{\nu \text{P}} }{dT} d\nu  }{\frac{dI_{\text{P}} }{dT} }  
\end{eqnarray} 
--- \textit{Rosseland} radiation mean free path, see \cite{Rosseland1924}; 
\begin{eqnarray} \label{I_P} 
I_{\text{P}} =\int _{0}^{\infty }I_{\nu \text{P}} d\nu  =\frac{\sigma T^{4} }{\pi} 
\end{eqnarray} 
--- total (obtained by integrating over all frequencies from $ 0 $ to $ \infty $) intensity of the equilibrium radiation;
\begin{eqnarray} \label{U_P} 
U_{\text{P}} =\int _{0}^{\infty }U_{\nu \text{P}} d\nu  =\frac{4\sigma T^{4} }{c} 
\end{eqnarray} 
--- total equilibrium radiant energy density; 
$\sigma $ --- Stefan-Boltzmann constant. 
The total equilibrium radiant energy density enters the gas-dynamic energy transfer equation through the equation of state of a substance with radiation.

The equation \eqref{S_RHC} corresponds to the second diffusion approximation equation \eqref{dif_approx_eq2} and describes well radiation propagation in optically thick media, but it is unsuitable for describing radiation propagation in optically thin media.
The use of geometric radiation mean free paths to reduce the value of the radiant energy flux, calculated by the formula \eqref{S_RHC}, which is anomalously large in optically thin media, is ill-founded.
Another critical weakness of the radiation heat conduction approximation is associated with the vertically breaking thermal wave profile, which is obtained in this approximation (see, for example, \cite[Chapter~X, \S~3]{Zeldovich2012}); as a result, an infinitely large value of radiant energy flux can be calculated at the thermal wave front using formula \eqref{S_RHC} by refining the grid in difference programs for calculating radiation propagation.

For optically thick media, the radiation heat conduction approximation has the zero order of the precision of the asymptotic approximation, since only the equilibrium radiant energy density enters the gas-dynamic energy transfer equation
\begin{eqnarray} \label{U_0} 
U_{\text{P}} = U^{\left(0\right)} ,
\end{eqnarray} 
usual boundary conditions on radiation of the zero order of the asymptotic precision are used, and the expression for radiation pressure is used that follows from an equation of state of a substance with radiation.

In calculations of radiation propagation in optically thick media at critical points, the limitations on radiation mean free paths can be used, which are similar to the equations \eqref{rep_l_t} and \eqref{rep_l_s}.

For optically thin media, we introduce a small parameter $\xi $ into the equation \eqref{kin_eq}:
\begin{eqnarray} \label{xi_thin} 
l^{\prime}_{\nu } \left(\frac{\partial }{c\partial t} +\mathbf{\Omega }\cdot \nabla \right)I_{\nu } =\xi \left(I_{\nu \text{P}} -I_{\nu } \right)\, , 
\end{eqnarray} 
\begin{eqnarray} \label{L_l_to_0} 
\frac{L}{l^{\prime}_{\nu } } \to 0\, ; 
\end{eqnarray} 
by writing the asymptotic expansion of the radiation intensity in the form of a formal series of successive approximations in powers of $\xi $
\begin{eqnarray} \label{exp_I_thin} 
I_{\nu } =\xi ^{0} I_{\nu }^{\left(0\right)} +\xi I_{\nu }^{\left(1\right)} +\xi ^{2} I_{\nu }^{\left(2\right)} +\cdots \, \, , 
\end{eqnarray} 
by substituting this power series into the equation \eqref{xi_thin} and equating the variable coefficients with the same powers $\xi $, we obtain the following system of equations of the method of successive approximations:

\begin{subequations}
	\label{eqn_I_thin_01n}
	\begin{eqnarray}
	l^{\prime}_{\nu } \left(\frac{\partial }{c\partial t} +\mathbf{\Omega }\cdot \nabla \right)I_{\nu }^{\left(0\right)}   &=&  0\, ,
	\label{eqI_0}
	\\
	l^{\prime}_{\nu } \left(\frac{\partial }{c\partial t} +\mathbf{\Omega }\cdot \nabla \right)I_{\nu }^{\left(1\right)}   &=&  \left(I_{\nu \text{P}} -I_{\nu }^{\left(0\right)} \right),
	\label{eqI_1}
	\\
	&{\vdots }& \nonumber\\
	l^{\prime}_{\nu } \left(\frac{\partial }{c\partial t} +\mathbf{\Omega }\cdot \nabla \right)I_{\nu }^{\left(n\right)}   &=&  -I_{\nu }^{\left(n-1\right)} .
	\label{eqI_n}
	\end{eqnarray}
\end{subequations}

Taking into account\eqref{def_U} and \eqref{def_S}, we obtain the system of equations for the asymptotic expansions of the spectral radiant energy density
\begin{eqnarray} \label{22)} 
U_{\nu } =\xi ^{0} U_{\nu }^{\left(0\right)} +\xi U_{\nu }^{\left(1\right)} +\xi ^{2} U_{\nu }^{\left(2\right)} +\cdots  
\end{eqnarray} 
and the spectral radiant energy flux
\begin{eqnarray} \label{23)} 
\mathbf{S}_{\nu } =\xi ^{0} \mathbf{S}_{\nu }^{\left(0\right)} +\xi \mathbf{S}_{\nu }^{\left(1\right)} +\xi ^{2} \mathbf{S}_{\nu }^{\left(2\right)} +\cdots \; , 
\end{eqnarray} 
from \eqref{exp_I_thin} and \eqref{eqn_I_thin_01n},
which can be written in the following form:
\begin{subequations}
	\label{eqn_U_thin_01n}
	\begin{eqnarray}
	\frac{\partial U_{\nu }^{\left(0\right)} }{\partial t} +\text{div}\, \mathbf{S}_{\nu }^{\left(0\right)}   &=&  0\, ,
	\label{eqU_0}
	\\
	\frac{\partial U_{\nu }^{\left(1\right)} }{\partial t} +\text{div}\, \mathbf{S}_{\nu }^{\left(1\right)}   &=&  c \kappa ^{\prime}_{\nu } \left(U_{\nu \text{P}} -U_{\nu }^{\left(0\right)} \right),
	\label{eqU_1}
	\\
	&{\vdots }& \nonumber\\
	\frac{\partial U_{\nu }^{\left(n\right)} }{\partial t} +\text{div}\, \mathbf{S}_{\nu }^{\left(n\right)}   &=&  -c \kappa ^{\prime}_{\nu } \, U_{\nu }^{\left(n-1\right)} .
	\label{eqU_n}
	\end{eqnarray}
\end{subequations}

Unfortunately, the system of equations \eqref{eqn_U_thin_01n} does not give us any useful results; 
after summing up the equations of system \eqref{eqn_U_thin_01n}, we simply obtain the first diffusion approximation equation \eqref{dif_approx_eq1}.

\section{Integral solution of the kinetic equation of radiation propagation}
\label{sec:integral_solution}

In the medium being in local thermodynamic equilibrium in the absence of radiation scattering leading to change of motion direction of the photons, we can consider the radiation propagation along the line defined by the equation
\begin{eqnarray}\label{line_eq}
\mathbf{r}\left( s \right)=\mathbf{r}_0 + s\, \mathbf{\Omega } ,	
\end{eqnarray}
and rewrite the equation \eqref{kin_eq} in the form:
\begin{eqnarray} \label{line_kin_eq} 
\left(\frac{\partial }{c\partial t} +\frac{\partial }{\partial s} \right)I_{\nu } =\kappa ^{\prime}_{\nu } \left(I_{\nu \text{P}} -I_{\nu } \right)\, . 
\end{eqnarray} 

The solution of the equation \eqref{line_kin_eq} is well known:
\begin{eqnarray} \label{line_int_sol} 
I_{\nu } \left(t,s,\mathbf{\Omega } \right)&=&\int _{s_b}^{s}\left(\kappa ^{\prime}_{\nu } I_{\nu \text{P}} \right)_{t-\frac{s-s^{\prime}}{c},\,  s^{\prime}} \exp \left[-\int _{s^{\prime}}^{s}\left(\kappa ^{\prime}_{\nu } \right)_{t\, -\, \frac{s-s^{\prime\prime}}{c} ,\, s^{\prime\prime}} d s^{\prime\prime} \right] d s^{\prime}+ \nonumber\\*
 &&+
\left(I_{\nu b} \right)_{t\, -\, \frac{s}{c},\, s_b} 
\exp \left[-\int_{s_b}^{s}\left(\kappa ^{\prime}_{\nu } 
\right)_{t\, -\, \frac{s-s^{\prime\prime}}{c},\, s^{\prime\prime} } d s^{\prime\prime} \right]\, , 
\end{eqnarray} 
cf., for example, the equation (2.33) in \cite{Zeldovich2012}.
$I_{\nu b}$ in \eqref{line_int_sol} denotes an arbitrary integration constant corresponding to the intensity of the entering radiation at the point 
$ \mathbf{r}\left( s_b \right) $ (or simply $s_b$, $ s_b \leq s $) 
on (left) border of the medium at the time $t-\frac{s}{c} $ in the direction defined by the unit vector $\mathbf{\Omega }$.
The solution \eqref{line_int_sol} may be verified by its direct substitution  into the equation \eqref{line_kin_eq}. 

According to the first summand in the solution \eqref{line_int_sol}, the radiation at the point $s$ of the considered line is formed by photons, ``born''  at points $s^{\prime}$ of the segment $\left[s_b,s\right]$ at earlier times, among which only the fraction  
\begin{eqnarray} \label{fraction} 
\exp \left[-\int _{s^{\prime}}^{s}\left(\kappa ^{\prime}_{\nu } \right)_{t-\frac{s-s^{\prime\prime}}{c},\, s^{\prime\prime} } d s^{\prime\prime} \right]
\end{eqnarray} 
arrives at the point $s$. 
Therefore, if 
\begin{eqnarray} \label{inf_line_cond} 
\kappa ^{\prime}_{\nu } \left( {s - s_b} \right) \gg 1 ,
\end{eqnarray} 
we can put $ {s_b} =  - \infty $ in the equation \eqref{line_int_sol} and obtain expression
\begin{eqnarray} \label{line_int_sol_inf} 
 &&I_{\nu } \left(t,\mathbf{r}_0,\mathbf{\Omega } \right)
= \int _{- \infty}^{0}\left(\kappa ^{\prime}_{\nu } I_{\nu \text{P}} \right)_{t+\frac{s^{\prime}}{c},\,  s^{\prime}} \exp \left[-\int _{s^{\prime}}^{0}\left(\kappa ^{\prime}_{\nu } \right)_{t\, +\, \frac{s^{\prime\prime}}{c} ,\, s^{\prime\prime}} d s^{\prime\prime} \right] d s^{\prime} \nonumber\\*
 &&=
\int _{0}^{\infty }I_{\nu \text{P}} e^{-\tau } d\tau 
=I_{\nu \text{P}} +\sum _{i=1}^{n}\left. \frac{\partial ^{i} I_{\nu \text{P}} }{\partial \tau ^{i} } \right| _{\tau =0} +\int _{0}^{\infty }\frac{\partial ^{n+1} I_{\nu \text{P}} }{\partial \tau ^{n+1} } e^{-\tau } d\tau  
\end{eqnarray} 
for the radiation intensity at the point $ \mathbf{r}_0 $ ($ s=0 $), where
\begin{subequations}
	\label{def_tau}
	\begin{eqnarray}
	\tau =\int _{s^{\prime}}^{0}\left(\kappa ^{\prime}_{\nu } \right)_{t\, +\, \frac{s^{\prime\prime}}{c} ,\, s^{\prime\prime}} d s^{\prime\prime}\, ,
	\label{def_tau_a}
	\\
	d\tau =-\left(\kappa ^{\prime}_{\nu } \right)_{t\, +\, \frac{s^{\prime}}{c} ,\, s^{\prime}} d s^{\prime} .
	\label{def_tau_b}
	\end{eqnarray}
\end{subequations}

According to the equations \eqref{line_eq}, \eqref{def_tau} and \eqref{I_n}
\begin{eqnarray} \label{dn_dtaun} 
\left. \frac{\partial ^{n} I_{\nu \text{P}} }{\partial \tau ^{n} } \right| _{\tau =0} 
&=& \left. \left [ \frac{\partial }{\partial \tau }
\frac{\partial ^{n-1} I_{\nu \text{P}} }{\partial \tau ^{n-1} }\right ] \right| _{\tau =0} 
= \left. \left [ \left(\frac{\partial s^{\prime}}{c\, \partial \tau} \frac{\partial }{\partial t} +\frac{\partial s^{\prime}}{\partial \tau}\frac{\partial }{\partial s^{\prime}} \right) 
\frac{\partial ^{n-1} I_{\nu \text{P}} }{\partial \tau ^{n-1} }\right ] \right| _{\tau =0} \nonumber\\*
&=&\left. \left [ -l^{\prime}_{\nu } \left( \frac{\partial }{c\,\partial t} +\frac{\partial }{\partial s^{\prime}} \right) 
\frac{\partial ^{n-1} I_{\nu \text{P}} }{\partial \tau ^{n-1} }\right ] \right| _{\tau =0} \nonumber\\* 
&=& \left. \left [ -l^{\prime}_{\nu } \left( \frac{\partial }{c\,\partial t} + \mathbf{\Omega }\cdot \nabla \right) 
\frac{\partial ^{n-1} I_{\nu \text{P}} }{\partial \tau ^{n-1} }\right ] \right| _{\tau =0}
= I_{\nu }^{\left(n\right)}.
\end{eqnarray} 
Thus, the asymptotic solution \eqref{exp_I_thick}, \eqref{exp_I_thick_01n} differs from the exact integral solution \eqref{line_int_sol_inf} only in the term 
\begin{eqnarray} \label{residual}
\int _{0}^{\infty }\frac{\partial ^{n+1} I_{\nu \text{P}} }{\partial \tau ^{n+1} } e^{-\tau } d\tau ,
\end{eqnarray}
which is  small for optically thick media and is tending to zero as $ \left( \frac{l^{\prime}_{\nu } }{L}\right)^{n+1} $, i.e. the asymptotic solution is asymptotic expansion of the exact integral solution, as it has to be.
Uncommon time derivatives in the equation \eqref{dn_dtaun} are connected with the time delay of radiation $ {t+\frac{s^{\prime}}{c}} $ in the equation \eqref{line_int_sol_inf}.

\section{Improved boundary conditions}
\label{sec:b_condition}

Proceeding to the derivation of improved boundary conditions, let us assume that the boundary between two substances can be locally considered plane --- this is a close approximation if the mean free path of photons is much less than the radius of curvature of the boundary
\begin{eqnarray} \label{plane_condition)} 
\frac{l^{\prime}_{\nu } }{R} \ll 1. 
\end{eqnarray} 
Let us choose spherical coordinate system $\left(r,\theta ,\varphi \right)$ in space with the origin on the boundary [$ \mathbf{r}_0 = 0 $, see the equation \eqref{line_eq}].
Let us assume that plane symmetry takes place, i.e. parameters of substances depend only on the coordinate 
\begin{eqnarray} \label{z_r_cos} 
z= s \cos \theta = r\cos \left(\pi -\theta \right) = - r \cos \theta 
\end{eqnarray} 
and time $t$. 

Let us consider the case when the medium, in which the radiation propagates, occupies the infinite half-space $z\le 0$ limited by the plane surface $z=0$  (the polar axis is directed along the outer normal to the surface $z=0$).
Taking into account the equation  \eqref{z_r_cos}, we obtain from the equations \eqref{line_int_sol_inf}, \eqref{dn_dtaun} the following expression for the spectral radiation intensity emerging from the surface of the substance occupying the left half-space $z\le 0$
\begin{eqnarray} \label{Im_nu} 
 &&I_{\nu }^{-} \left(t,0,\theta ,\varphi \right)
\,{\simeq}\, I_{\nu \text{P}}^{-} 
-\left. \cos \theta \left(l_{\nu }^{\prime -} \frac{\partial I_{\nu \text{P}}^{-} }{\partial z} \right)\right|_{z=-0}
-\left. \left(l_{\nu }^{\prime -} \frac{\partial I_{\nu \text{P}}^{-} }{c\partial t} \right)\right|_{z=-0}
\nonumber\\*
 &&\quad +\sum _{i=2}^{n}\left. 
\left(-1 \right)^i\left(\cos \theta \right)^i \underbrace{\left(l_{\nu }^{\prime -} \frac{\partial }{\partial z} \right)\cdots 
	\left(l_{\nu }^{\prime -} \frac{\partial }{\partial z} \right)}_{i-1} \left(l_{\nu }^{\prime -} \frac{\partial I_{\nu \text{P}}^{-} }{\partial z} \right)\right|_{z=-0} 
\nonumber\\*
 &&\quad +\sum _{i=2}^{n}\sum _{j=1}^{i-1}\left. 
\left(-1 \right)^i \left(\cos \theta \right)^{j} 
\left[ \cdots {\left(l_{\nu }^{\prime -} \frac{\partial }{\partial z} \right)} \left(l_{\nu }^{\prime -} \frac{\partial I_{\nu \text{P}}^{-} }{c\partial t} \right) \cdots + \cdots \right] \right|_{z=-0} 
\nonumber\\*
 &&\quad +\sum _{i=2}^{n}\sum _{j=1}^{i-1}\left. 
\left(-1 \right)^i \left(\cos \theta \right)^{j} 
\left[ \cdots {\left(l_{\nu }^{\prime -} \frac{\partial }{c\partial t} \right)} \left(l_{\nu }^{\prime -} \frac{\partial I_{\nu \text{P}}^{-} }{\partial z} \right) \cdots + \cdots \right] \right|_{z=-0} 
\nonumber\\*
 &&\quad +\sum _{i=2}^{n}\left. 
\left(-1 \right)^i \underbrace{\left(l_{\nu }^{\prime -} \frac{\partial }{c\partial t} \right)\cdots 
	\left(l_{\nu }^{\prime -} \frac{\partial }{c\partial t} \right)}_{i-1} \left(l_{\nu }^{\prime -} \frac{\partial I_{\nu \text{P}}^{-} }{c\partial t} \right)\right|_{z=-0} .
\end{eqnarray} 
All summands in the right part of the equation \eqref{Im_nu} relate to the boundary $z=-0$.

Taking into account the equations \eqref{def_U} and \eqref{U0_nu}, we obtain the expression for the spectral density of radiation in the left substance in the vicinity of the surface by integrating \eqref{Im_nu} over angles for the right half-sphere
\begin{eqnarray} \label{Um_nu} 
U_{\nu }^{-}  &{\simeq}&  \frac{1}{c} \int _{\left(2\pi \right)}I_{\nu }^{-} \left(t,0,\theta ,\varphi \right)d\mathbf{\Omega }\nonumber\\*
 &=&
\frac{1}{2} U_{\nu \text{P}}^{-} 
-\frac{1}{4} \left. \left(l_{\nu }^{\prime -} \frac{\partial U_{\nu \text{P}}^{-} }{\partial z} \right)\right|_{z=-0}
-\frac{1}{2} \left. \left(l_{\nu }^{\prime -} \frac{\partial U_{\nu \text{P}}^{-} }{c\partial t} \right)\right|_{z=-0}
\nonumber\\*
 && +\frac{1}{2}\sum _{i=2}^{n}\left. 
\frac{\left(-1 \right)^i}{i+1} 
\underbrace{\left(l_{\nu }^{\prime -} \frac{\partial }{\partial z} \right)\cdots 
	\left(l_{\nu }^{\prime -} \frac{\partial }{\partial z} \right)}_{i-1} \left(l_{\nu }^{\prime -} \frac{\partial U_{\nu \text{P}}^{-} }{\partial z} \right)\right|_{z=-0} 
\nonumber\\*
 && +\frac{1}{2}\sum _{i=2}^{n}\sum _{j=1}^{i-1}\left. 
\frac{\left(-1 \right)^i}{j+1} 
\left[ \cdots {\left(l_{\nu }^{\prime -} \frac{\partial }{\partial z} \right)} \left(l_{\nu }^{\prime -} \frac{\partial U_{\nu \text{P}}^{-} }{c\partial t} \right) \cdots + \cdots \right] \right|_{z=-0} 
\nonumber\\*
 && +\frac{1}{2}\sum _{i=2}^{n}\sum _{j=1}^{i-1}\left. 
\frac{\left(-1 \right)^i}{j+1} 
\left[ \cdots {\left(l_{\nu }^{\prime -} \frac{\partial }{c\partial t} \right)} \left(l_{\nu }^{\prime -} \frac{\partial U_{\nu \text{P}}^{-} }{\partial z} \right) \cdots + \cdots \right] \right|_{z=-0} 
\nonumber\\*
 && +\frac{1}{2}\sum _{i=2}^{n}\left. 
\left(-1 \right)^i 
\underbrace{\left(l_{\nu }^{\prime -} \frac{\partial }{c\partial t} \right)\cdots 
	\left(l_{\nu }^{\prime -} \frac{\partial }{c\partial t} \right)}_{i-1} \left(l_{\nu }^{\prime -} \frac{\partial U_{\nu \text{P}}^{-} }{c\partial t} \right)\right|_{z=-0} . 
\end{eqnarray} 
All summands in the right part of the equation \eqref{Um_nu} relate to the boundary $z=-0$.

Taking into account the equation \eqref{def_S}, the projection on the axis $z$ of the spectral radiant energy flux from the surface of the left substance is found by multiplying the equation \eqref{Im_nu} by   $\cos \theta $ and integrating over angles for the right half-sphere
\begin{eqnarray} \label{Sm_nu} 
S_{\nu }^{-}  &{\simeq}&  \int _{\left(2\pi \right)}I_{\nu }^{-} \left(t,0,\theta ,\varphi \right)\cos \theta \, d\mathbf{\Omega }
\nonumber\\*
 &=&\frac{c}{4} U_{\nu \text{P}}^{-} 
-\frac{c}{6} \left. \left(l_{\nu }^{\prime -} \frac{\partial U_{\nu \text{P}}^{-} }{\partial z} \right)\right|_{z=-0}
-\frac{c}{4} \left. \left(l_{\nu }^{\prime -} \frac{\partial U_{\nu \text{P}}^{-} }{c\partial t} \right)\right|_{z=-0}
\nonumber\\*
 && +\frac{c}{2}\sum _{i=2}^{n}\left. 
\frac{\left(-1 \right)^i}{i+2} 
\underbrace{\left(l_{\nu }^{\prime -} \frac{\partial }{\partial z} \right)\cdots 
	\left(l_{\nu }^{\prime -} \frac{\partial }{\partial z} \right)}_{i-1} \left(l_{\nu }^{\prime -} \frac{\partial U_{\nu \text{P}}^{-} }{\partial z} \right)\right|_{z=-0}  
\nonumber\\*
 && +\frac{c}{2}\sum _{i=2}^{n}\sum _{j=1}^{i-1}\left. 
\frac{\left(-1 \right)^i}{j+2} 
\left[ \cdots{\left(l_{\nu }^{\prime -} \frac{\partial }{\partial z} \right)} \left(l_{\nu }^{\prime -} \frac{\partial U_{\nu \text{P}}^{-} }{c\partial t} \right) \cdots + \cdots \right] \right|_{z=-0} 
\nonumber\\*
 && +\frac{c}{2}\sum _{i=2}^{n}\sum _{j=1}^{i-1}\left. 
\frac{\left(-1 \right)^i}{j+2} 
\left[ \cdots {\left(l_{\nu }^{\prime -} \frac{\partial }{c\partial t} \right)} \left(l_{\nu }^{\prime -} \frac{\partial U_{\nu \text{P}}^{-} }{\partial z} \right) \cdots + \cdots \right] \right|_{z=-0} 
\nonumber\\*
 && +\frac{c}{4}\sum _{i=2}^{n}\left. 
\left(-1 \right)^i 
\underbrace{\left(l_{\nu }^{\prime -} \frac{\partial }{c\partial t} \right)\cdots 
	\left(l_{\nu }^{\prime -} \frac{\partial }{c\partial t} \right)}_{i-1} \left(l_{\nu }^{\prime -} \frac{\partial U_{\nu \text{P}}^{-} }{c\partial t} \right)\right|_{z=-0} . 
\end{eqnarray} 
All summands in the right part of the equation \eqref{Sm_nu} relate to the boundary $z=-0$.

Comparing the expressions \eqref{Um_nu}, \eqref{Sm_nu} and confining ourselves in the right parts of the equations \eqref{Um_nu}, \eqref{Sm_nu} only to the first summands, we obtain, in particular, the relation
\begin{eqnarray} \label{diff_appr_b_cond} 
S_{\nu }^{-{\left(0\right)}} \approx \frac{c}{2} U_{\nu }^{-{\left(0\right)}}
= \frac{c}{4} U_{\nu \text{P}}^{-}\, , 
\end{eqnarray} 
which is often used as the boundary condition at the boundary between a substance and vacuum in the diffusion approximation, but the condition \eqref{Sm_nu} is \underbar{more accurate}, 
it takes into account a non-uniformity of substance characteristics in the direction, which is perpendicular to the substance surface. 

The flux of the $ z $-component of the momentum, which is transferred by light quanta  with momentum $ h\nu  \left/ c \right.$ through the surface with the normal directed along the axis $ z $, is determined by the $ {\text{p}}_{\nu }^{33} $-component of the spectral pressure tensor of radiation --- see the equation \eqref{def_P_nu_ab}. 
Multiplying the equation \eqref{Im_nu} by ${\cos ^2}\theta $ and integrating over angles for the right half-sphere, we obtain
\begin{eqnarray} \label{Pm_nu_zz} 
{\text{p}}_{\nu }^{33 -}  &{\simeq}&  \frac{1}{c} \int _{\left(2\pi \right)}I_{\nu }^{-} \left(t,0,\theta ,\varphi \right){\cos ^2}\theta  \,  d\mathbf{\Omega }
\nonumber\\*
 &=&\frac{1}{6} U_{\nu \text{P}}^{-} 
-\frac{1}{8} \left. \left(l_{\nu }^{\prime -} \frac{\partial U_{\nu \text{P}}^{-} }{\partial z} \right)\right|_{z=-0}
-\frac{1}{6} \left. \left(l_{\nu }^{\prime -} \frac{\partial U_{\nu \text{P}}^{-} }{c\partial t} \right)\right|_{z=-0}
\nonumber\\*
 && +\frac{1}{2}\sum _{i=2}^{n}\left. 
\frac{\left(-1 \right)^i}{i+3} 
\underbrace{\left(l_{\nu }^{\prime -} \frac{\partial }{\partial z} \right)\cdots 
	\left(l_{\nu }^{\prime -} \frac{\partial }{\partial z} \right)}_{i-1} \left(l_{\nu }^{\prime -} \frac{\partial U_{\nu \text{P}}^{-} }{\partial z} \right)\right|_{z=-0} 
\nonumber\\*
 && +\frac{1}{2}\sum _{i=2}^{n}\sum _{j=1}^{i-1}\left. 
\frac{\left(-1 \right)^i}{j+3} 
\left[ \cdots  {\left(l_{\nu }^{\prime -} \frac{\partial }{\partial z} \right)
} \left(l_{\nu }^{\prime -} \frac{\partial U_{\nu \text{P}}^{-} }{c\partial t} \right)\cdots  + \cdots \right] \right|_{z=-0} 
\nonumber\\*
 && +\frac{1}{2}\sum _{i=2}^{n}\sum _{j=1}^{i-1}\left. 
\frac{\left(-1 \right)^i}{j+3} 
\left[ \cdots  {\left(l_{\nu }^{\prime -} \frac{\partial }{c\partial t} \right)
} \left(l_{\nu }^{\prime -} \frac{\partial U_{\nu \text{P}}^{-} }{\partial z} \right)\cdots  + \cdots \right] \right|_{z=-0} 
\nonumber\\*
 && +\frac{1}{6}\sum _{i=2}^{n}\left. 
\left(-1 \right)^i 
\underbrace{\left(l_{\nu }^{\prime -} \frac{\partial }{c\partial t} \right)\cdots 
	\left(l_{\nu }^{\prime -} \frac{\partial }{c\partial t} \right)}_{i-1} \left(l_{\nu }^{\prime -} \frac{\partial U_{\nu \text{P}}^{-} }{c\partial t} \right)\right|_{z=-0} . 
\end{eqnarray} 
All summands in the right part of the equation \eqref{Pm_nu_zz} relate to the boundary $z=-0$.

Expressions, which are similar to the expressions \eqref{Im_nu}--\eqref{Sm_nu} and \eqref{Pm_nu_zz}, are derived for the substance occupying the right half-space $z\ge 0$ after integration over angles for the left half-sphere:
\begin{eqnarray} \label{Ip_nu} 
 && I_{\nu }^{+} \left(t,0,\theta ,\varphi \right)
\ {\simeq}\   I_{\nu \text{P}}^{+} 
-\left. \cos \theta \left(l_{\nu }^{\prime +} \frac{\partial I_{\nu \text{P}}^{+} }{\partial z} \right)\right|_{z=+0}
-\left. \left(l_{\nu }^{\prime +} \frac{\partial I_{\nu \text{P}}^{+} }{c\partial t} \right)\right|_{z=+0}
\nonumber\\*
 && \quad +\sum _{i=2}^{n}\left. 
\left(-1 \right)^i\left(\cos \theta \right)^i \underbrace{\left(l_{\nu }^{\prime +} \frac{\partial }{\partial z} \right)\cdots 
	\left(l_{\nu }^{\prime +} \frac{\partial }{\partial z} \right)}_{i-1} \left(l_{\nu }^{\prime +} \frac{\partial I_{\nu \text{P}}^{+} }{\partial z} \right)\right|_{z=+0} 
\nonumber\\*
 && \quad +\sum _{i=2}^{n}\sum _{j=1}^{i-1}\left. 
\left(-1 \right)^i \left(\cos \theta \right)^{j} 
\left[ \cdots {\left(l_{\nu }^{\prime +} \frac{\partial }{\partial z} \right)} \left(l_{\nu }^{\prime +} \frac{\partial I_{\nu \text{P}}^{+} }{c\partial t} \right) \cdots + \cdots \right] \right|_{z=+0} 
\nonumber\\*
 && \quad +\sum _{i=2}^{n}\sum _{j=1}^{i-1}\left. 
\left(-1 \right)^i \left(\cos \theta \right)^{j} 
\left[ \cdots {\left(l_{\nu }^{\prime +} \frac{\partial }{c\partial t} \right)} \left(l_{\nu }^{\prime +} \frac{\partial I_{\nu \text{P}}^{+} }{\partial z} \right) \cdots + \cdots \right] \right|_{z=+0} 
\nonumber\\*
 && \quad +\sum _{i=2}^{n}\left. 
\left(-1 \right)^i \underbrace{\left(l_{\nu }^{\prime +} \frac{\partial }{c\partial t} \right)\cdots 
	\left(l_{\nu }^{\prime +} \frac{\partial }{c\partial t} \right)}_{i-1} \left(l_{\nu }^{\prime +} \frac{\partial I_{\nu \text{P}}^{+} }{c\partial t} \right)\right|_{z=+0} ,
\end{eqnarray} 
\begin{eqnarray} \label{Up_nu} 
U_{\nu }^{+}  &{\simeq}&  \frac{1}{c} \int _{\left(2\pi \right)}I_{\nu }^{+} \left(t,0,\theta ,\varphi \right)d\mathbf{\Omega }\nonumber\\*
 &=&
\frac{1}{2} U_{\nu \text{P}}^{+} 
+\frac{1}{4} \left. \left(l_{\nu }^{\prime +} \frac{\partial U_{\nu \text{P}}^{+} }{\partial z} \right)\right|_{z=+0}
-\frac{1}{2} \left. \left(l_{\nu }^{\prime +} \frac{\partial U_{\nu \text{P}}^{+} }{c\partial t} \right)\right|_{z=+0}
\nonumber\\*
 && +\frac{1}{2}\sum _{i=2}^{n}\left. 
\frac{1}{i+1} 
\underbrace{\left(l_{\nu }^{\prime +} \frac{\partial }{\partial z} \right)\cdots 
	\left(l_{\nu }^{\prime +} \frac{\partial }{\partial z} \right)}_{i-1} \left(l_{\nu }^{\prime +} \frac{\partial U_{\nu \text{P}}^{+} }{\partial z} \right)\right|_{z=+0} 
\nonumber\\*
 && +\frac{1}{2}\sum _{i=2}^{n}\sum _{j=1}^{i-1}\left. 
\frac{\left(-1 \right)^{i+j}}{j+1} 
\left[ \cdots {\left(l_{\nu }^{\prime +} \frac{\partial }{\partial z} \right)} \left(l_{\nu }^{\prime +} \frac{\partial U_{\nu \text{P}}^{+} }{c\partial t} \right) \cdots + \cdots \right] \right|_{z=+0} 
\nonumber\\*
 && +\frac{1}{2}\sum _{i=2}^{n}\sum _{j=1}^{i-1}\left. 
\frac{\left(-1 \right)^{i+j}}{j+1} 
\left[ \cdots {\left(l_{\nu }^{\prime +} \frac{\partial }{c\partial t} \right)} \left(l_{\nu }^{\prime +} \frac{\partial U_{\nu \text{P}}^{+} }{\partial z} \right) \cdots + \cdots \right] \right|_{z=+0} 
\nonumber\\*
 && -\frac{1}{2}\sum _{i=2}^{n}\left. 
\underbrace{\left(l_{\nu }^{\prime +} \frac{\partial }{c\partial t} \right)\cdots 
	\left(l_{\nu }^{\prime +} \frac{\partial }{c\partial t} \right)}_{i-1} \left(l_{\nu }^{\prime +} \frac{\partial U_{\nu \text{P}}^{+} }{c\partial t} \right)\right|_{z=+0} , 
\end{eqnarray} 
\begin{eqnarray} \label{Sp_nu} 
S_{\nu }^{+}  &{\simeq}&  \int _{\left(2\pi \right)}I_{\nu }^{+} \left(t,0,\theta ,\varphi \right)\cos \theta \, d\mathbf{\Omega }
\nonumber\\*
 &=&-\frac{c}{4} U_{\nu \text{P}}^{+} 
-\frac{c}{6} \left. \left(l_{\nu }^{\prime +} \frac{\partial U_{\nu \text{P}}^{+} }{\partial z} \right)\right|_{z=+0}
+\frac{c}{4} \left. \left(l_{\nu }^{\prime +} \frac{\partial U_{\nu \text{P}}^{+} }{c\partial t} \right)\right|_{z=+0}
\nonumber\\*
 && -\frac{c}{2}\sum _{i=2}^{n}\left. 
\frac{1}{i+2} 
\underbrace{\left(l_{\nu }^{\prime +} \frac{\partial }{\partial z} \right)\cdots 
	\left(l_{\nu }^{\prime +} \frac{\partial }{\partial z} \right)}_{i-1} \left(l_{\nu }^{\prime +} \frac{\partial U_{\nu \text{P}}^{+} }{\partial z} \right)\right|_{z=+0}  
\nonumber\\*
 && -\frac{c}{2}\sum _{i=2}^{n}\sum _{j=1}^{i-1}\left. 
\frac{\left(-1 \right)^{i+j}}{j+2} 
\left[ \cdots {\left(l_{\nu }^{\prime +} \frac{\partial }{\partial z} \right)} \left(l_{\nu }^{\prime +} \frac{\partial U_{\nu \text{P}}^{+} }{c\partial t} \right) \cdots + \cdots \right] \right|_{z=+0} 
\nonumber\\*
 && -\frac{c}{2}\sum _{i=2}^{n}\sum _{j=1}^{i-1}\left. 
\frac{\left(-1 \right)^{i+j}}{j+2} 
\left[ \cdots {\left(l_{\nu }^{\prime +} \frac{\partial }{c\partial t} \right)} \left(l_{\nu }^{\prime +} \frac{\partial U_{\nu \text{P}}^{+} }{\partial z} \right) \cdots + \cdots \right] \right|_{z=+0} 
\nonumber\\*
 && -\frac{c}{4}\sum _{i=2}^{n}\left. 
\left(-1 \right)^i 
\underbrace{\left(l_{\nu }^{\prime +} \frac{\partial }{c\partial t} \right)\cdots 
	\left(l_{\nu }^{\prime +} \frac{\partial }{c\partial t} \right)}_{i-1} \left(l_{\nu }^{\prime +} \frac{\partial U_{\nu \text{P}}^{+} }{c\partial t} \right)\right|_{z=+0} , 
\end{eqnarray} 
\begin{eqnarray} \label{Pp_nu_zz} 
{\text{p}}_{\nu }^{33 +}  &{\simeq}&  \frac{1}{c} \int _{\left(2\pi \right)}I_{\nu }^{+} \left(t,0,\theta ,\varphi \right){\cos ^2}\theta  \,  d\mathbf{\Omega }
\nonumber\\*
 &=&	\frac{1}{6} U_{\nu \text{P}}^{+} 
+\frac{1}{8} \left. \left(l_{\nu }^{\prime +} \frac{\partial U_{\nu \text{P}}^{+} }{\partial z} \right)\right|_{z=+0}
-\frac{1}{6} \left. \left(l_{\nu }^{\prime +} \frac{\partial U_{\nu \text{P}}^{+} }{c\partial t} \right)\right|_{z=+0}
\nonumber\\*
 && +\frac{1}{2}\sum _{i=2}^{n}\left. 
\frac{1}{i+3} 
\underbrace{\left(l_{\nu }^{\prime +} \frac{\partial }{\partial z} \right)\cdots 
	\left(l_{\nu }^{\prime +} \frac{\partial }{\partial z} \right)}_{i-1} \left(l_{\nu }^{\prime +} \frac{\partial U_{\nu \text{P}}^{+} }{\partial z} \right)\right|_{z=+0} 
\nonumber\\*
 && +\frac{1}{2}\sum _{i=2}^{n}\sum _{j=1}^{i-1}\left. 
\frac{\left(-1 \right)^{i+j}}{j+3} 
\left[ \cdots {\left(l_{\nu }^{\prime +} \frac{\partial }{\partial z} \right)} \left(l_{\nu }^{\prime +} \frac{\partial U_{\nu \text{P}}^{+} }{c\partial t} \right) \cdots + \cdots \right] \right|_{z=+0} 
\nonumber\\*
 && +\frac{1}{2}\sum _{i=2}^{n}\sum _{j=1}^{i-1}\left. 
\frac{\left(-1 \right)^{i+j}}{j+3} 
\left[ \cdots {\left(l_{\nu }^{\prime +} \frac{\partial }{c\partial t} \right)} \left(l_{\nu }^{\prime +} \frac{\partial U_{\nu \text{P}}^{+} }{\partial z} \right) \cdots + \cdots \right] \right|_{z=+0} 
\nonumber\\*
 && -\frac{1}{6}\sum _{i=2}^{n}\left. 
\underbrace{\left(l_{\nu }^{\prime +} \frac{\partial }{c\partial t} \right)\cdots 
	\left(l_{\nu }^{\prime +} \frac{\partial }{c\partial t} \right)}_{i-1} \left(l_{\nu }^{\prime +} \frac{\partial U_{\nu \text{P}}^{+} }{c\partial t} \right)\right|_{z=+0} . 
\end{eqnarray} 
All summands in the right parts of the equations \eqref{Ip_nu}--\eqref{Pp_nu_zz} relate to the boundary $z=+0$.

The expressions \eqref{Um_nu}--\eqref{Sm_nu}, \eqref{Pm_nu_zz} and \eqref{Up_nu}--\eqref{Pp_nu_zz} can be used in calculations of radiation propagation only if they and the series \eqref{Im_nu}, \eqref{Ip_nu} converge. 
As in the section \ref{sec:asympt_solution}, the formal convergence of these series can be achieved due to limitations on radiation mean free path, which are similar to the equations \eqref{rep_l_t} and \eqref{rep_l_s}.

At the outer boundary of the substance with vacuum in the asymptotic approximation of the $n$th order and in the spectral diffusion approximation with increased to the $n$th order precision (see suggestion 2 in the Conclusions below), it is proposed to use following expressions for normal components of the spectral radiant energy flux as boundary conditions, cf. the equations \eqref{Sm_nu} and \eqref{Sp_nu}:
\begin{eqnarray} \label{tildaSm_nu} 
\tilde{S}_{\nu }^{-} 
 &=& \frac{c}{4} U_{\nu \text{P}}^{-} 
-\frac{c}{6} \left. \left(\tilde{l}_{\nu }^{\,\prime -} \frac{\partial U_{\nu \text{P}}^{-} }{\partial z} \right)\right|_{z=-0}
-\frac{c}{4} \left. \left(\tilde{l}_{\nu }^{\,\prime -} \frac{\partial U_{\nu \text{P}}^{-} }{c\partial t} \right)\right|_{z=-0}
\nonumber\\*
 && +\frac{c}{2}\sum _{i=2}^{n}\left. 
\frac{\left(-1 \right)^i}{i+2} 
\underbrace{\left(\tilde{l}_{\nu }^{\,\prime -} \frac{\partial }{\partial z} \right)\cdots 
	\left(\tilde{l}_{\nu }^{\,\prime -} \frac{\partial }{\partial z} \right)}_{i-1} \left(\tilde{l}_{\nu }^{\,\prime -} \frac{\partial U_{\nu \text{P}}^{-} }{\partial z} \right)\right|_{z=-0}  
\nonumber\\*
 && +\frac{c}{2}\sum _{i=2}^{n}\sum _{j=1}^{i-1}\left. 
\frac{\left(-1 \right)^i}{j+2} 
\left[ \cdots {\left(\tilde{l}_{\nu }^{\,\prime -} \frac{\partial }{\partial z} \right)} \left(\tilde{l}_{\nu }^{\,\prime -} \frac{\partial U_{\nu \text{P}}^{-} }{c\partial t} \right) \cdots + \cdots \right] \right|_{z=-0} 
\nonumber\\*
 && +\frac{c}{2}\sum _{i=2}^{n}\sum _{j=1}^{i-1}\left. 
\frac{\left(-1 \right)^i}{j+2} 
\left[ \cdots {\left(\tilde{l}_{\nu }^{\,\prime -} \frac{\partial }{c\partial t} \right)} \left(\tilde{l}_{\nu }^{\,\prime -} \frac{\partial U_{\nu \text{P}}^{-} }{\partial z} \right) \cdots + \cdots \right] \right|_{z=-0} 
\nonumber\\*
 && +\frac{c}{4}\sum _{i=2}^{n}\left. 
\left(-1 \right)^i 
\underbrace{\left(\tilde{l}_{\nu }^{\,\prime -} \frac{\partial }{c\partial t} \right)\cdots 
	\left(\tilde{l}_{\nu }^{\,\prime -} \frac{\partial }{c\partial t} \right)}_{i-1} \left(\tilde{l}_{\nu }^{\,\prime -} \frac{\partial U_{\nu \text{P}}^{-} }{c\partial t} \right)\right|_{z=-0} , 
\end{eqnarray} 
\begin{eqnarray} \label{tildaSp_nu} 
\tilde{S}_{\nu }^{+}
 &=& -\frac{c}{4} U_{\nu \text{P}}^{+} 
-\frac{c}{6} \left. \left(\tilde{l}_{\nu }^{\,\prime +} \frac{\partial U_{\nu \text{P}}^{+} }{\partial z} \right)\right|_{z=+0}
+\frac{c}{4} \left. \left(\tilde{l}_{\nu }^{\,\prime +} \frac{\partial U_{\nu \text{P}}^{+} }{c\partial t} \right)\right|_{z=+0}
\nonumber\\*
 && -\frac{c}{2}\sum _{i=2}^{n}\left. 
\frac{1}{i+2} 
\underbrace{\left(\tilde{l}_{\nu }^{\,\prime +} \frac{\partial }{\partial z} \right)\cdots 
	\left(\tilde{l}_{\nu }^{\,\prime +} \frac{\partial }{\partial z} \right)}_{i-1} \left(\tilde{l}_{\nu }^{\,\prime +} \frac{\partial U_{\nu \text{P}}^{+} }{\partial z} \right)\right|_{z=+0}  
\nonumber\\*
 && -\frac{c}{2}\sum _{i=2}^{n}\sum _{j=1}^{i-1}\left. 
\frac{\left(-1 \right)^{i+j}}{j+2} 
\left[ \cdots {\left(\tilde{l}_{\nu }^{\,\prime +} \frac{\partial }{\partial z} \right)} \left(\tilde{l}_{\nu }^{\,\prime +} \frac{\partial U_{\nu \text{P}}^{+} }{c\partial t} \right) \cdots + \cdots \right] \right|_{z=+0} 
\nonumber\\*
 && -\frac{c}{2}\sum _{i=2}^{n}\sum _{j=1}^{i-1}\left. 
\frac{\left(-1 \right)^{i+j}}{j+2} 
\left[ \cdots {\left(\tilde{l}_{\nu }^{\,\prime +} \frac{\partial }{c\partial t} \right)} \left(\tilde{l}_{\nu }^{\,\prime +} \frac{\partial U_{\nu \text{P}}^{+} }{\partial z} \right) \cdots + \cdots \right] \right|_{z=+0} 
\nonumber\\*
 && -\frac{c}{4}\sum _{i=2}^{n}\left. 
\left(-1 \right)^i 
\underbrace{\left(\tilde{l}_{\nu }^{\,\prime +} \frac{\partial }{c\partial t} \right)\cdots 
	\left(\tilde{l}_{\nu }^{\,\prime +} \frac{\partial }{c\partial t} \right)}_{i-1} \left(\tilde{l}_{\nu }^{\,\prime +} \frac{\partial U_{\nu \text{P}}^{+} }{c\partial t} \right)\right|_{z=+0} , 
\end{eqnarray} 
all summands in the right parts of the equations \eqref{tildaSm_nu}, \eqref{tildaSp_nu} are calculated at the substance boundary at points $z=-0$ and $z=+0$, respectively.

The projection of the spectral radiant energy flux on the normal to the boundary of two substances is equal to the algebraic sum of $\tilde{S}_{\nu }^{-} $ and $\tilde{S}_{\nu }^{+} $ flux projections:
\begin{eqnarray} \label{Sb_nu} 
\tilde{S}_{\nu }^{\text{b}} 
 &=&\tilde{S}_{\nu }^{-} +\tilde{S}_{\nu }^{+} 
\nonumber\\*
 &=& \frac{c}{4} \left(U_{\nu \text{P}}^{-} -U_{\nu \text{P}}^{+} \right)
-\frac{c}{6} \left[\left. \left(\tilde{l}_{\nu }^{\,\prime -} \frac{\partial U_{\nu \text{P}}^{-} }{\partial z} \right)\right|_{z=-0} +\left. \left(\tilde{l}_{\nu }^{\,\prime +} \frac{\partial U_{\nu \text{P}}^{+} }{\partial z} \right)\right|_{z=+0}\, \right]  
\nonumber\\*
 && -\frac{c}{4} \left[ \left. \left(\tilde{l}_{\nu }^{\,\prime -} \frac{\partial U_{\nu \text{P}}^{-} }{c\partial t} \right)\right|_{z=-0} 
- \left. \left(\tilde{l}_{\nu }^{\,\prime +} \frac{\partial U_{\nu \text{P}}^{+} }{c\partial t} \right)\right|_{z=+0}\, \right] + \ldots . 
\end{eqnarray} 
It is proposed to use expression \eqref{Sb_nu} 
at the inner boundaries in the asymptotic approximation of the $n$th order and in the spectral diffusion approximation with the precision increased to the $n$th order. 
If one equate $U_{\nu \text{P}}^{-} =U_{\nu \text{P}}^{+} $ and partial time derivatives in the equation \eqref{Sb_nu}, the ability to correctly calculate radiation propagation in problems with initial data discontinuity will be lost.

At the outer boundaries with vacuum for normal components of the total (integrated over all frequencies) radiant energy flux in the approximation of radiation heat conduction, we propose to use expressions
\begin{eqnarray} \label{tildaSm} 
\tilde{S}^{-} 
= \frac{c}{4} U_{\text{P}}^{-} 
-\frac{c}{6} \left. \left(\tilde{l}_\text{R}^{\,\prime -} \frac{\partial U_{\text{P}}^{-} }{\partial z} \right)\right|_{z=-0}
-\frac{c}{4} \left. \left(\tilde{l}_\text{R}^{\,\prime -} \frac{\partial U_{\text{P}}^{-} }{c\partial t} \right)\right|_{z=-0} ,
\end{eqnarray} 
\begin{eqnarray} \label{tildaSp} 
\tilde{S}^{+}
= -\frac{c}{4} U_{\text{P}}^{+} 
-\frac{c}{6} \left. \left(\tilde{l}_\text{R}^{\,\prime +} \frac{\partial U_{\text{P}}^{+} }{\partial z} \right)\right|_{z=+0}
+\frac{c}{4} \left. \left(\tilde{l}_\text{R}^{\,\prime +} \frac{\partial U_{\text{P}}^{+} }{c\partial t} \right)\right|_{z=+0} , 
\end{eqnarray} 
corresponding to the expressions \eqref{tildaSm_nu} and \eqref{tildaSp_nu}, 
with limitations similar to the limitations \eqref{rep_l_t} and \eqref{rep_l_s}. 

We propose to use an expression corresponding to the equation \eqref{Sb_nu} for the normal component of the total radiant energy flux
\begin{eqnarray} \label{S_b} 
\tilde{S}^{\text{b}} 
 &=&\tilde{S}^{-} +\tilde{S}^{+} 
\nonumber\\*
 &=&
\frac{c}{4} \left(U_{\text{P}}^{-} -U_{\text{P}}^{+} \right)
-\frac{c}{6} \left[\left. \left(\tilde{l}_\text{R}^{\,\prime -} \frac{\partial U_{\text{P}}^{-} }{\partial z} \right)\right|_{z=-0} +\left. \left(\tilde{l}_\text{R}^{\,\prime +} \frac{\partial U_{\text{P}}^{+} }{\partial z} \right)\right|_{z=+0}\, \right]  
\nonumber\\*
 && -\frac{c}{4} \left[ \left. \left(\tilde{l}_\text{R}^{\,\prime -} \frac{\partial U_{\text{P}}^{-} }{c\partial t} \right)\right|_{z=-0} 
- \left. \left(\tilde{l}_\text{R}^{\,\prime +} \frac{\partial U_{\text{P}}^{+} }{c\partial t} \right)\right|_{z=+0}\, \right] ,
\end{eqnarray} 
with limitations similar to the limitations \eqref{rep_l_t} and \eqref{rep_l_s}, at inner boundaries in the approximation of radiation heat conduction. 
As well as in the equation \eqref{Sb_nu}, if one equate $U_{\text{P}}^{-} =U_{\text{P}}^{+} $ and partial time derivatives in the equation \eqref{S_b}, then the ability to correctly calculate radiation propagation in problems with initial data discontinuity will be lost.

\section{Conclusions}

The $n$th order asymptotic approximation obtained from the asymptotic solution of the kinetic equation of radiation propagation can be used as follows:
\begin{enumerate}
	\item to replace the diffusion approximation or the radiation heat conduction approximation with the $n$th order asymptotic approximation;
	\item to increase the asymptotic precision of gas-dynamic calculations with radiation (without scattering) in the diffusion approximation to the $n$th order by using the expression \eqref{exp_S_thick} for the spectral radiant energy flux and the expression 
	\eqref{exp_P_nu_ab} for radiation pressure;
	\item to increase the precision of gas-dynamic calculations with radiation (without scattering) in the radiation heat conduction approximation to the first order by adding the radiant energy density of the first order
	\begin{eqnarray} \label{U_1}
	U^{\left(1\right)} =-\, \tilde{l}_\text{R}^{\,\prime} \frac{\partial U_{\text{P}} }{c\partial t} 
	\end{eqnarray}
	into the energy transfer equation, and the first order radiation pressure tensor
	\begin{eqnarray} \label{P_ab1} 
	{\text{p}}^{\alpha \beta \left(1\right)} 
	=\frac{1}{3} \left( - \tilde{l}_\text{R}^{\,\prime} \frac{\partial U_{\text{P}}}{c\partial t} \right) {g^{\alpha \beta}}
	= \frac{1}{3} U^{\left(1\right)} {g^{\alpha \beta}} , 
	\end{eqnarray} 
	into the energy transfer equation and into the momentum transfer equation;
dependence of radiation intensity on the angle appears in the first order of the approximation precision [see the equation \eqref{I_1}], nevertheless, due to vertically breaking profile of thermal waves in the radiation heat conduction approximation, 
it is necessary to use systematically limitations on radiation mean free paths similar to the equations \eqref{rep_l_t} and \eqref{rep_l_s} on fronts of thermal waves, what forces to doubt the possibility of the correct description of angular distribution of radiation intensity in the radiation heat conduction approximation as the result of reduction of the approximation precision in the general case.
\end{enumerate}

Boundary conditions of proper order of the asymptotic precision shall be used in calculations of radiation propagation 
when the new $n$th order asymptotic approximation or the diffusion approximation and the radiation heat conduction approximation of increased precision are used.

\bibliographystyle{apsrev4-1}
\bibliography{ssheat_er}

\begin{thebibliography}{13}%
\makeatletter
\providecommand \@ifxundefined [1]{%
 \@ifx{#1\undefined}
}%
\providecommand \@ifnum [1]{%
 \ifnum #1\expandafter \@firstoftwo
 \else \expandafter \@secondoftwo
 \fi
}%
\providecommand \@ifx [1]{%
 \ifx #1\expandafter \@firstoftwo
 \else \expandafter \@secondoftwo
 \fi
}%
\providecommand \natexlab [1]{#1}%
\providecommand \enquote  [1]{``#1''}%
\providecommand \bibnamefont  [1]{#1}%
\providecommand \bibfnamefont [1]{#1}%
\providecommand \citenamefont [1]{#1}%
\providecommand \href@noop [0]{\@secondoftwo}%
\providecommand \href [0]{\begingroup \@sanitize@url \@href}%
\providecommand \@href[1]{\@@startlink{#1}\@@href}%
\providecommand \@@href[1]{\endgroup#1\@@endlink}%
\providecommand \@sanitize@url [0]{\catcode `\\12\catcode `\$12\catcode
  `\&12\catcode `\#12\catcode `\^12\catcode `\_12\catcode `\%12\relax}%
\providecommand \@@startlink[1]{}%
\providecommand \@@endlink[0]{}%
\providecommand \url  [0]{\begingroup\@sanitize@url \@url }%
\providecommand \@url [1]{\endgroup\@href {#1}{\urlprefix }}%
\providecommand \urlprefix  [0]{URL }%
\providecommand \Eprint [0]{\href }%
\providecommand \doibase [0]{http://dx.doi.org/}%
\providecommand \selectlanguage [0]{\@gobble}%
\providecommand \bibinfo  [0]{\@secondoftwo}%
\providecommand \bibfield  [0]{\@secondoftwo}%
\providecommand \translation [1]{[#1]}%
\providecommand \BibitemOpen [0]{}%
\providecommand \bibitemStop [0]{}%
\providecommand \bibitemNoStop [0]{.\EOS\space}%
\providecommand \EOS [0]{\spacefactor3000\relax}%
\providecommand \BibitemShut  [1]{\csname bibitem#1\endcsname}%
\let\auto@bib@innerbib\@empty
\bibitem [{\citenamefont {Zel'dovich}\ and\ \citenamefont
  {Raizer}(2012)}]{Zeldovich2012}%
  \BibitemOpen
  \bibfield  {author} {\bibinfo {author} {\bibfnamefont {Y.~B.}\ \bibnamefont
  {Zel'dovich}}\ and\ \bibinfo {author} {\bibfnamefont {Y.~P.}\ \bibnamefont
  {Raizer}},\ }\href@noop {} {\emph {\bibinfo {title} {Physics of Shock Waves
  and High-Temperature Hydrodynamic Phenomena}}}\ (\bibinfo  {publisher} {Dover
  Publications Inc.},\ \bibinfo {address} {Mineola, New York},\ \bibinfo {year}
  {2012})\BibitemShut {NoStop}%
\bibitem [{\citenamefont {Mihalas}(1978)}]{Mihalas1978stellar}%
  \BibitemOpen
  \bibfield  {author} {\bibinfo {author} {\bibfnamefont {D.}~\bibnamefont
  {Mihalas}},\ }\href@noop {} {\emph {\bibinfo {title} {Stellar atmospheres}}}\
  (\bibinfo  {publisher} {W.~H.~Freeman and Co.},\ \bibinfo {address} {San
  Francisco},\ \bibinfo {year} {1978})\BibitemShut {NoStop}%
\bibitem [{\citenamefont {Larsen}\ \emph {et~al.}(1983)\citenamefont {Larsen},
  \citenamefont {Pomranin},\ and\ \citenamefont {Badham}}]{Larsen1983}%
  \BibitemOpen
  \bibfield  {author} {\bibinfo {author} {\bibfnamefont {E.~W.}\ \bibnamefont
  {Larsen}}, \bibinfo {author} {\bibfnamefont {G.~C.}\ \bibnamefont
  {Pomranin}}, \ and\ \bibinfo {author} {\bibfnamefont {V.~C.}\ \bibnamefont
  {Badham}},\ }\href@noop {} {\bibfield  {journal} {\bibinfo  {journal}
  {Journal of Quantitative Spectroscopy and Radiative Transfer}\ }\textbf
  {\bibinfo {volume} {29}},\ \bibinfo {pages} {285} (\bibinfo {year}
  {1983})}\BibitemShut {NoStop}%
\bibitem [{\citenamefont {Habetler}\ and\ \citenamefont
  {Matkowsky}(1975)}]{habetler1975uniform}%
  \BibitemOpen
  \bibfield  {author} {\bibinfo {author} {\bibfnamefont {G.~J.}\ \bibnamefont
  {Habetler}}\ and\ \bibinfo {author} {\bibfnamefont {B.~J.}\ \bibnamefont
  {Matkowsky}},\ }\href@noop {} {\bibfield  {journal} {\bibinfo  {journal}
  {Journal of Mathematical Physics}\ }\textbf {\bibinfo {volume} {16}},\
  \bibinfo {pages} {846} (\bibinfo {year} {1975})}\BibitemShut {NoStop}%
\bibitem [{\citenamefont {Larsen}\ and\ \citenamefont
  {D'Arruda}(1976)}]{larsen1976asymptotic}%
  \BibitemOpen
  \bibfield  {author} {\bibinfo {author} {\bibfnamefont {E.~W.}\ \bibnamefont
  {Larsen}}\ and\ \bibinfo {author} {\bibfnamefont {J.}~\bibnamefont
  {D'Arruda}},\ }\href@noop {} {\bibfield  {journal} {\bibinfo  {journal}
  {Physical Review A}\ }\textbf {\bibinfo {volume} {13}},\ \bibinfo {pages}
  {1933} (\bibinfo {year} {1976})}\BibitemShut {NoStop}%
\bibitem [{\citenamefont {Larsen}(1977)}]{larsen1977asymptotic}%
  \BibitemOpen
  \bibfield  {author} {\bibinfo {author} {\bibfnamefont {E.~W.}\ \bibnamefont
  {Larsen}},\ }\href@noop {} {\bibfield  {journal} {\bibinfo  {journal} {SIAM
  Journal on Applied Mathematics}\ }\textbf {\bibinfo {volume} {33}},\ \bibinfo
  {pages} {427} (\bibinfo {year} {1977})}\BibitemShut {NoStop}%
\bibitem [{\citenamefont {Larsen}(2010)}]{larsen2010asymptotic}%
  \BibitemOpen
  \bibfield  {author} {\bibinfo {author} {\bibfnamefont {E.~W.}\ \bibnamefont
  {Larsen}},\ }\href@noop {} {\bibfield  {journal} {\bibinfo  {journal}
  {Transport Theory and Statistical Physics}\ }\textbf {\bibinfo {volume}
  {39}},\ \bibinfo {pages} {110} (\bibinfo {year} {2010})}\BibitemShut
  {NoStop}%
\bibitem [{\citenamefont {Brunner}(2002)}]{Brunner2002forms}%
  \BibitemOpen
  \bibfield  {author} {\bibinfo {author} {\bibfnamefont {T.~A.}\ \bibnamefont
  {Brunner}},\ }\href@noop {} {\emph {\bibinfo {title} {Forms of approximate
  radiation transport}}},\ \bibinfo {type} {Sandia Report}\ (\bibinfo
  {institution} {Sandia National Laboratories},\ \bibinfo {year}
  {2002})\BibitemShut {NoStop}%
\bibitem [{\citenamefont {Ambartsumian}\ \emph {et~al.}(1958)\citenamefont
  {Ambartsumian} \emph {et~al.}}]{Ambartsumian1958}%
  \BibitemOpen
  \bibfield  {author} {\bibinfo {author} {\bibfnamefont {V.~A.}\ \bibnamefont
  {Ambartsumian}} \emph {et~al.},\ }\href@noop {} {\emph {\bibinfo {title}
  {Theoretical astrophysics}}}\ (\bibinfo  {publisher} {Pergamon Press},\
  \bibinfo {address} {New York},\ \bibinfo {year} {1958})\BibitemShut {NoStop}%
\bibitem [{\citenamefont {Chandrasekhar}(1960)}]{chandrasekhar1960radiative}%
  \BibitemOpen
  \bibfield  {author} {\bibinfo {author} {\bibfnamefont {S.}~\bibnamefont
  {Chandrasekhar}},\ }\href@noop {} {\emph {\bibinfo {title} {Radiative
  transfer}}}\ (\bibinfo  {publisher} {Dover Publications Inc.},\ \bibinfo
  {address} {New York},\ \bibinfo {year} {1960})\BibitemShut {NoStop}%
\bibitem [{\citenamefont {Bourbaki}(2004)}]{bourbaki2004}%
  \BibitemOpen
  \bibfield  {author} {\bibinfo {author} {\bibfnamefont {N.}~\bibnamefont
  {Bourbaki}},\ }\href@noop {} {\emph {\bibinfo {title} {Functions of a Real
  Variable}}},\ Elements of Mathematics, Book IV\ (\bibinfo  {publisher}
  {Springer},\ \bibinfo {address} {Berlin},\ \bibinfo {year}
  {2004})\BibitemShut {NoStop}%
\bibitem [{\citenamefont {Mihalas}\ and\ \citenamefont
  {Mihalas}(1995)}]{mihalas1995foundations}%
  \BibitemOpen
  \bibfield  {author} {\bibinfo {author} {\bibfnamefont {D.}~\bibnamefont
  {Mihalas}}\ and\ \bibinfo {author} {\bibfnamefont {B.~W.}\ \bibnamefont
  {Mihalas}},\ }\href@noop {} {\emph {\bibinfo {title} {Foundations of
  radiation hydrodynamics}}}\ (\bibinfo  {publisher} {Oxford University
  Press},\ \bibinfo {year} {1995})\BibitemShut {NoStop}%
\bibitem [{\citenamefont {Rosseland}(1965)}]{Rosseland1924}%
  \BibitemOpen
  \bibfield  {author} {\bibinfo {author} {\bibfnamefont {S.}~\bibnamefont
  {Rosseland}},\ }in\ \href@noop {} {\emph {\bibinfo {booktitle} {Selected
  Papers on the Transfer of Radiation}}},\ \bibinfo {editor} {edited by\
  \bibinfo {editor} {\bibfnamefont {D.~H.}\ \bibnamefont {Menzel}}}\ (\bibinfo
  {publisher} {Dover Publications Inc.},\ \bibinfo {address} {New York},\
  \bibinfo {year} {1965})\ pp.\ \bibinfo {pages} {73--76}\BibitemShut {NoStop}%
\end{thebibliography}%
	
\end{document}